\RequirePackage{fix-cm}
\documentclass[twocolumn,epjc3]{svjour3}  
\smartqed  

\usepackage{amsfonts}
\usepackage{amssymb}
\usepackage{amsmath}
\usepackage{cite}
\usepackage{cancel}
\RequirePackage{graphicx}
\usepackage{dcolumn}
\usepackage{bm}
\usepackage{slashed}
\usepackage{lineno}
\usepackage{IEEEtrantools}
\usepackage{csquotes}
\usepackage{makecell}
\usepackage{multirow}
\usepackage{cancel}
\usepackage{verbatim}
\usepackage{tikz-feynman}
\usepackage{siunitx}
\usepackage{graphicx} 
\usepackage{subcaption}
\usepackage{float}
\usepackage{placeins}
\usepackage[colorlinks = true,
            linkcolor = blue,
            urlcolor  = blue,
            citecolor = blue,
            anchorcolor = blue]{hyperref}
\usepackage{tikz,xcolor}
            
\DeclareOldFontCommand{\tt}{\normalfont\ttfamily}{\mathtt}
\definecolor{lime}{HTML}{A6CE39}
\DeclareRobustCommand{\orcidicon}{
	\begin{tikzpicture}
	\draw[lime, fill=lime] (0,0) 
	circle [radius=0.16] 
	node[white] {{\fontfamily{qag}\selectfont \tiny ID}};
	\draw[white, fill=white] (-0.0625,0.095) 
	circle [radius=0.007];
	\end{tikzpicture}
	\hspace{-2mm}
}

\foreach \x in {A, ..., Z}{\expandafter\xdef\csname orcid\x\endcsname{\noexpand\href{https://orcid.org/\csname orcidauthor\x\endcsname}
			{\noexpand\orcidicon}}
}

\journalname{Noname}
\begin{document}

\title{Phenomenology of photons-enriched semi-visible jets 
}


\author{Cesare Cazzaniga\thanksref{e1,addr1}\orcidB{}  
\and
        Alessandro Russo\thanksref{e3,addr2}\orcidD{} 
        \and
        Emre Sitti\thanksref{e4,addr1}\orcidE{}
           \and 
        Annapaola de Cosa\thanksref{e2,addr1}\orcidC{} 
}


\thankstext{e1}{e-mail: cesare.cazzaniga@cern.ch}

\thankstext{e2}{e-mail: adecosa@phys.ethz.ch}
\thankstext{e3}{e-mail: arusso00@stanford.edu}
\thankstext{e4}{e-mail: esitti@ethz.ch}

\institute{ 
ETH Z\"urich, Institute for Particle Physics and Astrophysics, CH-8093 Z\"urich, Switzerland 
\label{addr1}
\and 
\text{ }Stanford University, Department of Physics, Stanford, CA 94305, USA
\label{addr2}
}

\date{Received: date / Accepted: date}

\maketitle

\begin{abstract}
This Letter proposes a new signature for confining dark sectors at the LHC. 
Under the assumption of a QCD-like hidden sector, hadronic jets containing stable dark bound states originating from hidden strong dynamics, known as semi-visible jets, could manifest in proton-proton collisions. In the proposed simplified model, a heavy $Z'$ mediator coupling to SM quarks allows the resonant production of dark quarks, subsequently hadronizing in stable and unstable dark bound states. The unstable dark bound states can then decay back to SM quarks via the same $Z'$ portal or photons via a lighter pseudo-scalar portal (such as an axion-like particle). This mechanism creates a new signature where semi-visible jets are enriched in non-isolated photons. We show that these exotic jets evade the phase space probed by current LHC searches exploiting jets or photons due to the expected high jet neutral electromagnetic fraction and photons candidates non-isolation, respectively. In the proposed analysis strategy to tackle such signature, we exploit jets as final state objects to represent the underlying QCD-like hidden sector. We show that, by removing any selection on the neutral electromagnetic fraction from the jet identification criteria, higher signal efficiency can be reached. To enhance the signal-to-background discrimination, we train a deep neural network as a jet tagger that exploits differences in the substructure of signal and background jets. We estimate that with the available triggers for Run 2 and this new strategy, a high mass search can claim a $5 \sigma$ discovery (exclusion) of the $Z'$ boson with a mass up to \SI{5}{TeV} (\SI{5}{TeV}) with the full Run~2 data of the LHC when the fraction of unstable dark hadrons decaying to photons pairs is around $\SI{30}{\%}$, and with a coupling of the $Z'$ to SM quarks of 0.25.

\end{abstract}


\section{Introduction}
\label{intro}
The presence of dark matter (DM) in the universe~\cite{Bertone:2004pz,Planck2018} is clear evidence of limitations of the Standard Model (SM) framework, invoking new physics to find a full explanation of the phenomenon.  The lack of experimental confirmation of the most popular hypotheses investigated so far~\cite{PhysRevLett.39.165,Jungman:1995df} connects to the idea of new physics belonging to complex sectors of matter barely accessible to SM particles, \emph{Hidden sectors}. Hidden sectors are introduced in the Hidden Valley models~\cite{Strassler:2006im} as extensions of the SM, and connected to the SM via a heavy mediator or a weak coupling. These models can rather easily accommodate DM candidates, identifying them with stable particles within the dark sector~\cite{SpierMoreiraAlves:2010err,Lee:2015gsa,Okawa:2016wrr,Hochberg:2018vdo,Beauchesne:2018myj,Bernreuther:2019pfb,Beauchesne:2019ato}. Moreover, if realised in nature, the Hidden Valley scenario may result in unusual and little-studied phenomena at the LHC. 
A notable example is \emph{semi-visible jets (SVJ)} signatures appearing in strongly coupled Hidden Valley scenarios, where dark QCD exists and behaves in a QCD-like way.  Previous works~\cite{Cohen:2015toa,Cohen:2017pzm,Beauchesne:2017yhh,Beauchesne:2018myj,Bernreuther:2019pfb,Beauchesne:2019ato, Cazzaniga:2022hxl,Beauchesne_2023}, have introduced such class of experimental signatures, composed of a mixture of stable and unstable dark bound states originating from the hadronization of dark quarks at colliders. 
SVJs appear as partially visible jets of particles in the detectors, with missing transverse momentum from stable dark mesons ($\cancel{E}_{\text{T}}$) aligned to the direction of one of the jets. 
Fully hadronic SVJ signatures have been proposed in ~\cite{Cohen:2015toa,Cohen:2017pzm,Beauchesne:2017yhh,Bernreuther:2019pfb} and searched for by the CMS and ATLAS experiments in s- and t-channel production modes~\cite{CMS:2021dzg,ATLAS-CONF-2022-038}, respectively. 
Novel signatures with SVJ enriched in leptons from dark photon decays have been recently proposed in~\cite{Cazzaniga:2022hxl}, along with signatures enriched in $\tau$ leptons~\cite{Beauchesne_2023} proposed by models assuming non-democratic decays to leptons). 
In this Letter we present a simplified model for SVJs enriched in photons. The model has been developed following~\cite{Fox:2011qd,Knapen:2021eip,Cazzaniga:2022hxl,Beauchesne_2023}. In the proposed models, a $Z'$ mediator acts as a messenger between the visible and dark sectors via coupling to SM quarks, thus allowing the resonant production of dark quarks in proton-proton collisions and the decays of the unstable dark pions to SM quarks. In such way the SVJs can be resonantly produced at the LHC. The decay of the dark pions to photons is guaranteed via a pseudo-scalar (such as an axion-like particle, ALP) portal coupling to dark quarks and SM photons. This mechanism allows the production of non-isolated photon pairs inside semi-visible jets (SVJ$\gamma$~signature). As an extreme case of this signature, we also show that within our model, it is possible to obtain jets comprised exclusively of photons (photon-jets~\cite{Ellis:2012zp}). We find that present searches have limited sensitivity to these novel signatures since SVJ$\gamma$ objects do not pass standard identification criteria both for jets and photons. We propose a dedicated strategy to identify such objects with respect to jets faking photons and prompt photon production.

\section{Model setup}
\label{sec:Model}
\begin{figure}
\centering
\begin{tikzpicture}
    \begin{feynman}
      \vertex (a);
      \vertex[above left=0.75cm of a] (b1) {$q$};
      \vertex[below left=0.75cm of a] (b2) {$\Bar{q}$};
      \vertex[right=0.75cm of a] (c);
      \vertex[right=1.5cm of c] (d);
      \vertex[right=0.9cm of c](cd);
      \vertex[above=0.55cm of cd](cd2);
      \vertex[below=0.55cm of cd](cd4);
      
      \vertex[above=0.5cm of d] (d1);
      \vertex[above=0.5cm of d1] (d2);
      \vertex[below=0.5cm of d] (d3);
      \vertex[below=0.5cm of d3] (d4);
      \vertex[right=1.05cm of d] (e) {$\rho_v$};
      \vertex[above=0.75cm of e] (e1);
      \vertex[above=0.75cm of e1] (e2){$\rho_v$};
      \vertex[below=0.75cm of e] (e3);
      \vertex[below=0.75cm of e3] (e4){$\pi_v$};

      \vertex[right=0.75cm of e1] (f2);
      \vertex[right=0.75cm of f2] (g2);
      \vertex[above=0.2cm of g2] (ga) {$\gamma$};
      \vertex[below=0.2cm of g2] (gb) {$\gamma$};
  
      \vertex[right=0.75cm of e3] (f3);
      \vertex[right=0.75cm of f3] (g3);
      \vertex[above=0.2cm of g3] (gc) {$q$};
      \vertex[below=0.2cm of g3] (gd) {$\Bar{q}$};

      \diagram* {
        (a) -- [fermion] (b2),
        (b1) -- [fermion] (a),
        (a) -- [boson, edge label=$Z'$] (c),
        (c) -- [fermion, edge label=$\psi^v$] (d2),
        (cd2) -- [gluon] (d),
        (d4) -- [fermion, edge label=$\Bar{\psi^v}$] (c),
        (cd4) -- [gluon] (d),
        (d) -- (e),
        (d1) -- [edge label=$\pi_v$, pos=0.8] (e1),
        (d2) -- (e2),
        (d3) -- [edge label=$\pi_v$, pos=0.32] (e3),
        (d4) -- (e4),

        (e1) -- [scalar, edge label=$a$] (f2),
        (e3) -- [boson, edge label=$Z'$] (f3),

        (f2) -- [boson] (ga),
        (f2) -- [boson] (gb),

        (f3) -- [fermion] (gc),
        (gd) -- [fermion] (f3),
        
      };
      \filldraw[color=darkgray, fill=lightgray, ultra thick](d) ellipse (0.3cm and 1.1cm);
    \end{feynman}
\end{tikzpicture}
\caption{Feynman diagram illustrating the main processes in a collider experiment. The gray ellipses represent the hadronization part.}
\label{fig:FD}
\end{figure}
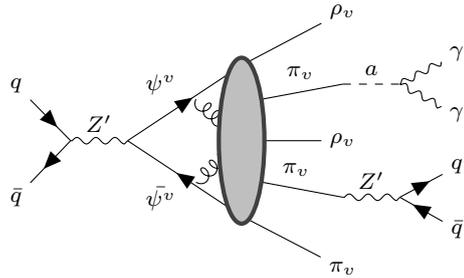

\begin{figure}
    \centering
    \includegraphics[width=1\linewidth]{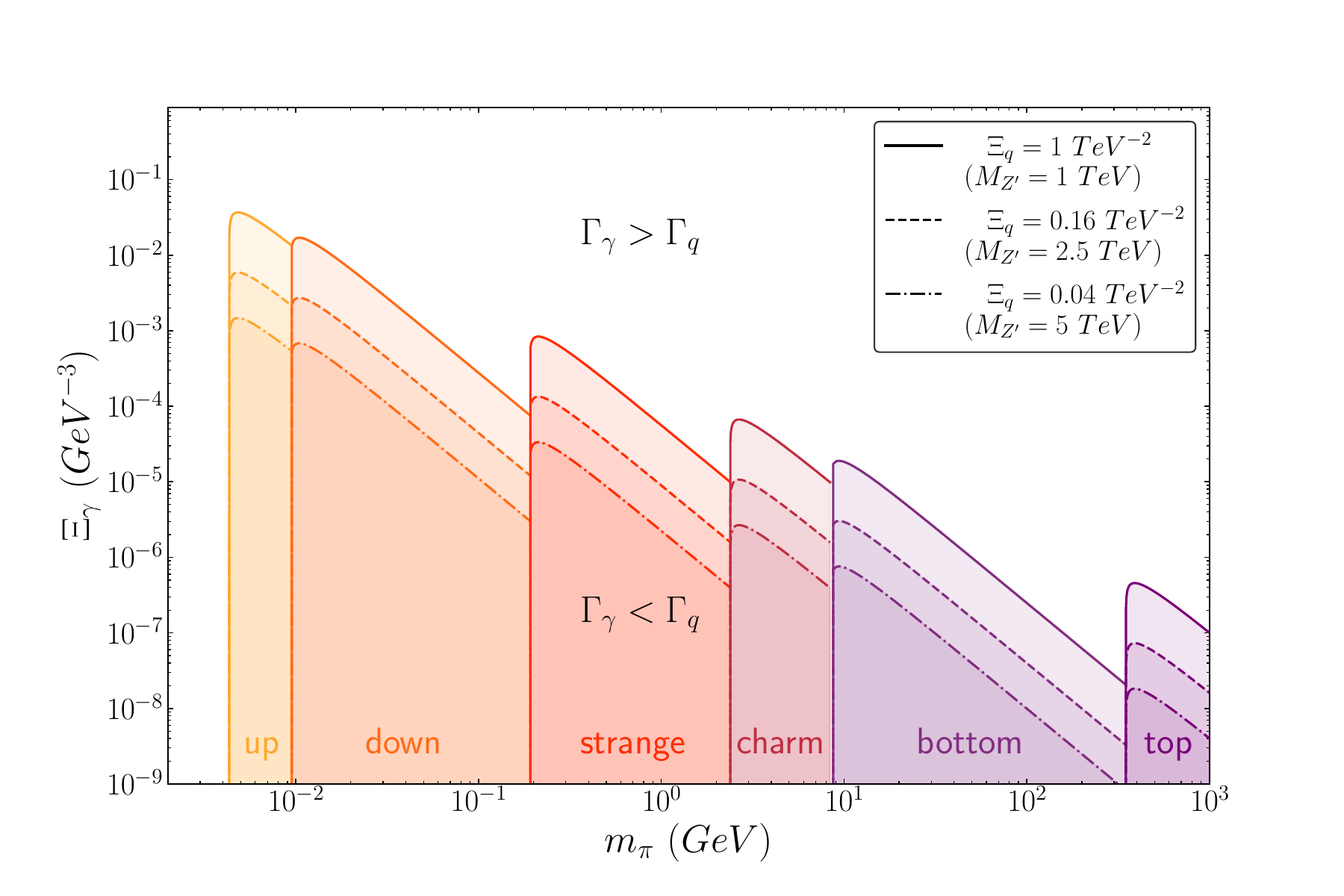}
    \caption{The plot shows seven different regions in parameter space ($\Xi_\gamma$ vs $m_{\pi_v}$) where different decay modes of the bound states are dominant. In the shaded regions below the lines, $\pi_v$ will predominantly decay into two SM quarks: up quarks in the yellow region, down quarks in the orange region, strange quarks in the red region, charm quarks in the cardinal region, bottom quarks in the violet region, and top quarks in the purple region. In the white region above the lines, the dominant decay mode of $\pi_v$ is into two photons. The lines themselves represent the case where the branching ratio between $\pi_v \to \gamma\gamma$ and $\pi_v \to \bar{q}q$ is exactly \SI{50}{\%}. This analysis has been performed for different values of $\Xi_q$.}
    \label{fig:BF_50p}
\end{figure}

The model is primarily based on the frameworks introduced in ~\cite{Fox:2011qd,Knapen:2021eip,Cazzaniga:2022hxl,Beauchesne_2023}, allowing the dark sector to communicate with the SM. As mentioned above, it employs a $Z'$ mediator and a pseudo-scalar ALP $a$. A simple description and summary of the physical processes are shown in Figure \ref{fig:FD}, and the main interactions are governed by the following Lagrangians:
\begin{equation}
\begin{split}
\mathcal{L}_{Z'} \supset & -Z{\mu}^{'} \bar{q_i}\gamma^{\mu}(g_{ij}^R P_R + g_{ij}^{L} P_L)q_j \\
& - Z_{\mu}^{'} \bar{\psi_{i}}\gamma^{\mu}(y_{ij}^{R} P_R + y_{ij}^{L} P_L)\psi_{j} \, \, ,
\end{split}
\label{eq:lag_Zp}
\end{equation}
for $Z'$, and
\begin{equation}
\begin{split}
\mathcal{L}_a \supset & \, (D_\mu a)^\dagger (D^\mu a) - \frac{1}{2}m_a^2 a^2 \\
& - i\lambda^L a \bar{L} \gamma_5 L
- i\lambda_{ij}^\psi a \bar{\psi}_i^v \gamma_5 \psi_j^v + \text{h.c.}
\end{split}
\label{eq:lag_a}
\end{equation}
for $a$.
Here, $q_i$ are the SM quarks of flavor $i$, while $\psi_i^v$ denotes the dark quarks. In what follows, we will omit the dark sector flavor indices to simplify the notation; however, this does not imply that there is only one flavor. We also assume the couplings of the $Z'$ to leptons to be negligible compared to quarks as suggested by the constraints from the high mass di-lepton searches~\cite{CMS:2018ipm,ATLAS:2019erb}. Additionally,  a vector-like lepton ($L$)~\cite{Knapen:2021eip} mediates the decay of the dark bound state into two photons, resulting in the following effective interaction:
\begin{equation}
\mathcal{L}_{\text{eff}} \supset \frac{\alpha_{EM} \, \lambda_L \lambda_\psi \Lambda_v F_{\pi_v}}{4\pi}\frac{\pi_v}{m_a^2 M_L}F_{\mu \nu} \tilde{F}^{\mu \nu} \, \, ,
\label{eq:lag_eff}
\end{equation}
where $\alpha_{EM}$ is the QED coupling, $M_L$ is the vector-like lepton mass, $\Lambda_v$ is the dark sector confinement scale, $F_{\pi_v}$ is the dark (pseudo-scalar) meson decay constant, $\pi_v$ is the lightest dark bound state and $F^{\mu \nu}$ is the electromagnetic field tensor. In addition, as discussed in \cite{Knapen:2021eip}, it is also possible to implement a gluon portal for the ALP, similar to the one for Eq. (\ref{eq:lag_eff}), but employing a vector-like quark $Q$ (of mass $M_Q$) instead of $L$. This new portal will, however, be suppressed by a factor of $(1/M_Q)^2$ in the decay width. Vector-like quarks have been extensively studied~\cite{CMS:2022fck} and are more constrained on the full range of masses with respect to vector-like leptons~\cite{CMS:2019hsm}. For this reason, this portal is neglected in favour of the photon channel.  
In a collider experiment, as shown in Figure~\ref{fig:FD}
, the dark sector is accessed through the $Z'$, which is initially produced by a parton interaction. Then the dark quarks hadronize, producing multiple hadrons. These states can decay into stable DM or back into the SM using the portals described in Eq. (\ref{eq:lag_Zp}) and (\ref{eq:lag_a}).\\ 
In the context of our model, there are two main hadrons with different dark isospin structures: $\rho$ and $\pi_v$, the latter being the lightest. Each of these states can have three different configurations depending on the dark sector flavors: $(\pi_v/\rho_v)^+$, $(\pi_v/\rho_v)^0$, and $(\pi_v/\rho_v)^-$. This should not be understood as an electromagnetic charge, as it arises from an accidental symmetry (isospin) which can be easily broken. For this reason, we will omit this notation and simply refer to the bound states as $\pi_v$ and $\rho_v$. Depending on the model's symmetry, the latter will decay into its lighter counterpart ($\pi_v$), which can then decay into either two SM quarks via the $Z'$ or two photons via the pseudo-scalar $a$. These are the ``visible" decays. However, it is also possible for the light mesons (or a fraction of them) to remain stable, making them potential DM candidates. The decay widths for the visible decays are as follows: 
\begin{equation}
    \Gamma(\pi_v \to \Bar{q_i}q_i)= \Gamma_q = \frac{N_c}{32 \pi} m_{q_i}^2 m_{\pi_v}^3 \sqrt{1-\frac{4m_{q_i}^2}{m_{\pi_v}^2}} \Xi_q ^2 \, \, , 
    \label{eq:gamma_q}
\end{equation}
\begin{equation}
    \Gamma(\pi_v \to \gamma \gamma) = \Gamma_\gamma= \frac{\alpha_{EM}^2}{64 \pi^2}m_{\pi_v}^7 \Xi_{\gamma}^2 \, \, .
    \label{eq:gamma_p}
\end{equation}
where $N_c$ is the number of SM colors, $m_{q_i}$ is the SM quark mass and $m_{\pi_v}$ is the dark pion mass. Here $\Xi_q$ and $\Xi_\gamma$ functions are effective couplings for the theory, keeping within themself some of the free degrees of freedom. They are defined as:
\begin{equation}
    \Xi_q := \frac{\Delta^q_i \Delta^{\psi}}{M_{Z'}^2} \, \, \, \text{ and } \, \, \,\Xi_{\gamma}:= \frac{\lambda_L\lambda_{\psi}}{m_a^2 M_L} \quad ,
    \label{eq:Xis}
\end{equation}
where $\Delta^q_i$ and $\Delta^\psi$ are, respectively,  $g_{ii}^{R} - g_{ii}^{L}$ and $y^{R}-y^{L}$, and the assumption $F_{\pi_v} \approx \Lambda_v \approx m_{\pi_v}$ was made.
Depending on the values of $\Xi_q$ and $\Xi_\gamma$, the bound states will decay back into photons or quarks. By studying the branching fraction in the latter case, one can observe that only one flavor of quark becomes relevant for the same value of $\Xi_q$ and given $m_{\pi_v}$. Specifically, if we order the quarks by their masses ($i<j \Rightarrow m_{q_i}<m_{q_j}$) and we have that $m_{q_i} < m_{\pi_v} < m_{q_{i+1}}$, then the flavor $i+1$ is forbidden, and the flavor $i$ dominates over $i-1$ by a factor of $(m_{q_i}/m_{q_{i-1}})^2$.\\
It is also possible to tune $\Xi_q$ and $\Xi_\gamma$ to achieve a specific value of $R_{\pi_v} := \Gamma_q/\Gamma_\gamma$. By adjusting this ratio, the number of photons or quarks coming from the bound states in the SVJ$\gamma$ can be controlled. Figure \ref{fig:BF_50p} shows the boundary between two scenarios as a function of the bound state mass and $\Xi_\gamma$. The shaded region below the lines represents the case where the bound states decay predominantly into quarks over photons while, above the lines, the opposite case occurs. This analysis is conducted for different values of $\Xi_q'$ (or $M_{Z'}$ by keeping the $\Delta$s unitary).
It is worth mentioning that in the specific case where $\Delta^q_i = 0$, so for a pure $Z'$ vector portal, the decays of the dark pions back to SM quarks via the $Z'$ are prohibited. In this specific scenario, the pseudo-scalar can still decay to photons via the ALP, leading to sprays of collimated photons resembling the photon-jets signature~\cite{Ellis:2012zp}, but with a potentially much larger expected number of photons pairs per jet.
In this study, we only consider dark-bound states to be heavy enough to decay promptly. However, longer lifetimes can be obtained, allowing for signatures with additional displaced tracks and vertices inside the signal jets and non-prompt, non-isolated photons.

\section{Monte Carlo Simulations}
\label{sec:MC simulations}
Samples containing signal events, including initial- and final-state radiation and underlying event, have been produced with {\tt Pythia8.307}~\cite{Bierlich:2022pfr}. 
The Hidden Valley module~\cite{Carloni:2011kk,Carloni:2010tw} in {\tt Pythia8} simulated the dark sector's showering, hadronization, and decays. 
Detector effects are simulated using {\tt Delphes3}~\cite{deFavereau:2013fsa}.
Following previous studies~\cite{CMS:2021dzg}, anti-$k_T$ jets~\cite{Cacciari:2008gp,Cacciari:2011ma} with \mbox{$R = 0.8$} (AK8 jets) are clustered requiring a minimum $p_{\text{T}}$ of \SI{200}{GeV}. 
All the samples have been normalised to the LO cross-section prediction computed with the {\tt MadGraph5\_aMC@NLO}~\cite{Alwall:2014hca} event generator using the {\tt NN23LO1} parton distribution functions ~\cite{Ball:2013hta} from the {\tt Lhapdf}~\cite{Buckley:2014ana} repository. As a reference, we have normalized the signals to the expected cross-section for $p p \to Z' \to q_v \bar{q}$ as in~\cite{CMS:2021dzg}, thus assuming a coupling of the $Z'$ to SM quarks equal to 0.25, and its coupling to dark quarks equal to 0.4, following the most recent recommendations from the LHC
DM Working Group~\cite{Boveia:2016mrp}. \\ \\
For the SVJ${\gamma}$ s-channel signal process, $20 \cdot 10^3$ events have been generated scanning a mass range for $M_{Z'}$ between \SI{1.5}{TeV} and \SI{5}{TeV}. 
The number of stable and unstable dark hadrons produced in the dark hadronization process can vary according to the details of the dark sector. 
In previous literature\cite{Cohen:2015toa,Cohen:2017pzm}, an effective invisible fraction parameter has been defined as $r_{\text{inv}}=  \langle {N_{\text{stable}}}/{(N_{\text{stable}} + N_{\text{unstable}})} \rangle$, where  $N_{\text{stable}}$ is the number of stable dark hadrons, and $N_{\text{unstable}}$ is the number of those that are unstable, decaying back to SM quarks and photons. This invisible fraction allows the capture of variations in the details of the hidden sector and defines the parameter space of a SVJ signature in terms of the amount of missing transverse momentum $\cancel{E}_{\text{T}}$.  The quantity $\cancel{E}_{\text{T}}$ is computed from the transverse component of the sum of the momenta of all detected particles in an event~\cite{deFavereau:2013fsa}, and can be used to quantify the amount of transverse momentum missing due to invisible particles.
The higher $r_{\text{inv}}$, the more the final state topology will resemble a mono-jet signature, where $\cancel{E}_{\text{T}}$ recoils against initial-state radiation jets.  
The dark pseudo-scalar ($\pi_v$) and vector ($\rho_v$) meson masses can differ according to the non-perturbative dynamics of the hidden sector, even for mass-degenerate dark quarks. 
Following a similar approach as in \cite{Cazzaniga:2022hxl,Beauchesne_2023}, we employed the lattice QCD fits in~\cite{Albouy:2022cin} to predict the masses of dark vector mesons $m_{\rho_v}$ from the input ratio $m_{\pi_v}/\Lambda_v$, where $\Lambda_v$ ({\tt HiddenValley:Lambda}) is the dark confinement scale, which fixes the overall mass scale for the dark bound states.  
As a benchmark, $\Lambda_v$ has been set to \SI{20}{GeV} and the ratio $m_{\pi_v}/\Lambda_v = 1.0$, therefore fixing $m_{\pi_v}$ ({\tt 4900111:m0, 4900211:m0}) at \SI{20}{GeV} and $m_{\rho_v}$ ({\tt 4900113:m0, 4900213:m0}) at $\sim$ \SI{54}{GeV}. 
With these dark QCD parameter settings, the $\rho_v \to \pi_v \pi_v$ decay is open, and we assume $100 \%$ branching ratio for this internal decay in the dark sector. 
Consequently, only unstable pseudo-scalar dark mesons are allowed to decay back to the SM within the chosen parameter space. As discussed in the previous section, the unstable bound states can decay via the $Z'$ portal to SM quarks and via the ALP to photons. The fraction of decays of the dark pseudo-scalars to photons can be tuned via the effective parameter $\text{BR}_{\gamma}$. This parameter can assume values between 0 and 1, and can be tuned by changing $R_{\pi_v}$ in the simplified model introduced in the previous section. \\
In summary, there are four parameters sensitive to the details of the dark sector: the confinement scale $\Lambda_v$, the pseudo-scalar mass ratio $m_{\pi_v}/\Lambda_v$, the invisible fraction $r_{\text{inv}}$, and the branching ratio to photons $\text{BR}_{\gamma}$. 
Moreover, there are three portal parameters related to the production of SVJ$\gamma$ in proton-proton collisions: the $Z'$ pole mass $M_{Z'}$, the coupling of the $Z'$ to dark quarks $g_{q_v}$ and the coupling to SM quarks $g_q$. The portal parameters related to the VLL and ALP are entered in the $\text{BR}_{\gamma}$ definition and are only relevant to control the bound states' decays to photons. 
In order to probe the impact of enhanced photon content, we scan over the branching $\text{BR}_{\gamma}~(0.3, 0.5, 0.7, 1)$ which is equivalent to varying the parameters $\Xi_q$  and $\Xi_p$. 
Similar to hadronic and leptonic SVJ signatures, SM QCD interactions, top pair productions, and electroweak processes are background sources for  SVJ$\gamma$. Moreover, prompt photons production ($\gamma + \text{jets}$) can also emulate signal jets. We, therefore, include this additional process in our studies.
All the background samples have been generated at LO with the {\tt MadGraph5\_aMC@NLO} event generator using the {\tt NN23LO1} parton distribution functions from the {\tt Lhapdf} repository. The evolution of the parton-level events and hadronization are performed with {\tt Pythia8}. 
The QCD sample ($4.5 \cdot 10^7$ events) has been produced requiring a generator level cut on the leading parton jet transverse momentum $p_{\text{T}} > \SI{500}{GeV}$. 
The QCD background is particularly relevant due to the large cross-section and the possibility of mis-reconstruction of the jet momentum, leading to additional missing momentum aligned with the jet axis. Moreover, QCD jets can contain photons from neutral SM pions decays as well as final state QED radiation, which might resemble the SVJ$\gamma$ signal.
The $\text{t} \bar{\text{t}} + \text{jets}$ inclusive sample ($5 \cdot 10^7$ events) has been generated with up to two additional partons. 
This background mainly becomes relevant when the top quarks are highly boosted, and therefore, the W boson decay and the b-initiated jets are merged into a larger jet. The electroweak inclusive backgrounds $\text{Z}(\nu \bar{\nu}) + \text{jets} $ and $\text{W} (\ell \nu) + \text{jets}$ ($2.5 \cdot 10^7$ events each) have been produced with a generator level cut $H_{\text{T}} > \SI{100}{GeV}$ and including up to three additional partons in the matrix element.
Finally, the $\gamma + \text{jets}$ inclusive sample ($2 \cdot 10^7$ events) has been generated with up to two additional partons and requiring a generator level cut on the leading parton jet transverse momentum $p_{\text{T}} > \SI{500}{GeV}$. This background is especially relevant when the expected fraction of photons in the signal jets is particularly high, and thus, the signal jet becomes particularly hard to distinguish from a prompt photon.

\begin{figure*} 
\centering 
\begin{subfigure}{.4\textwidth} 
\centering 
\includegraphics[width=.9\linewidth]{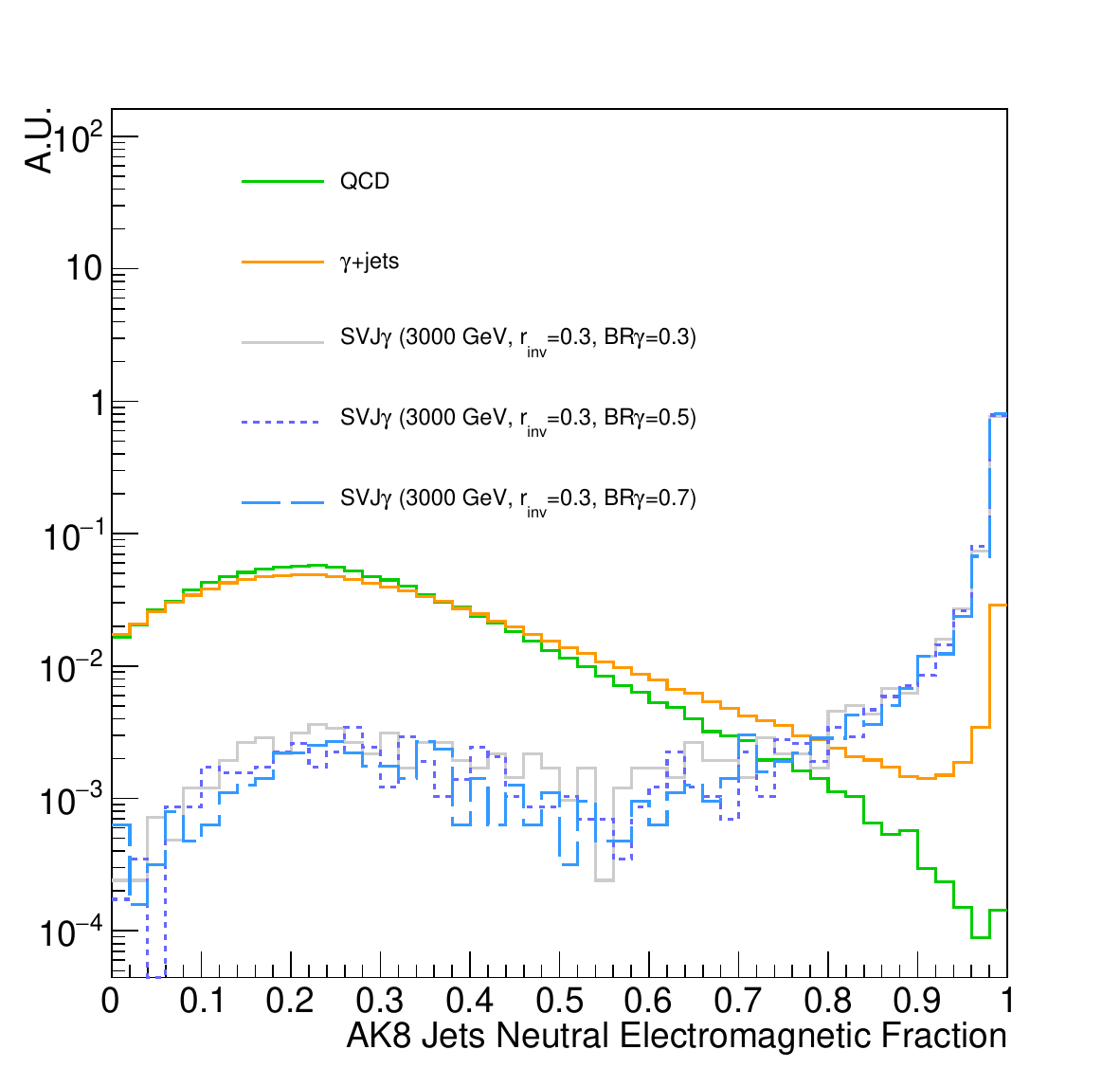} \label{fig:distr_nemf} 
\end{subfigure} 
\begin{subfigure}{.4\textwidth} \centering \includegraphics[width=.9\linewidth]{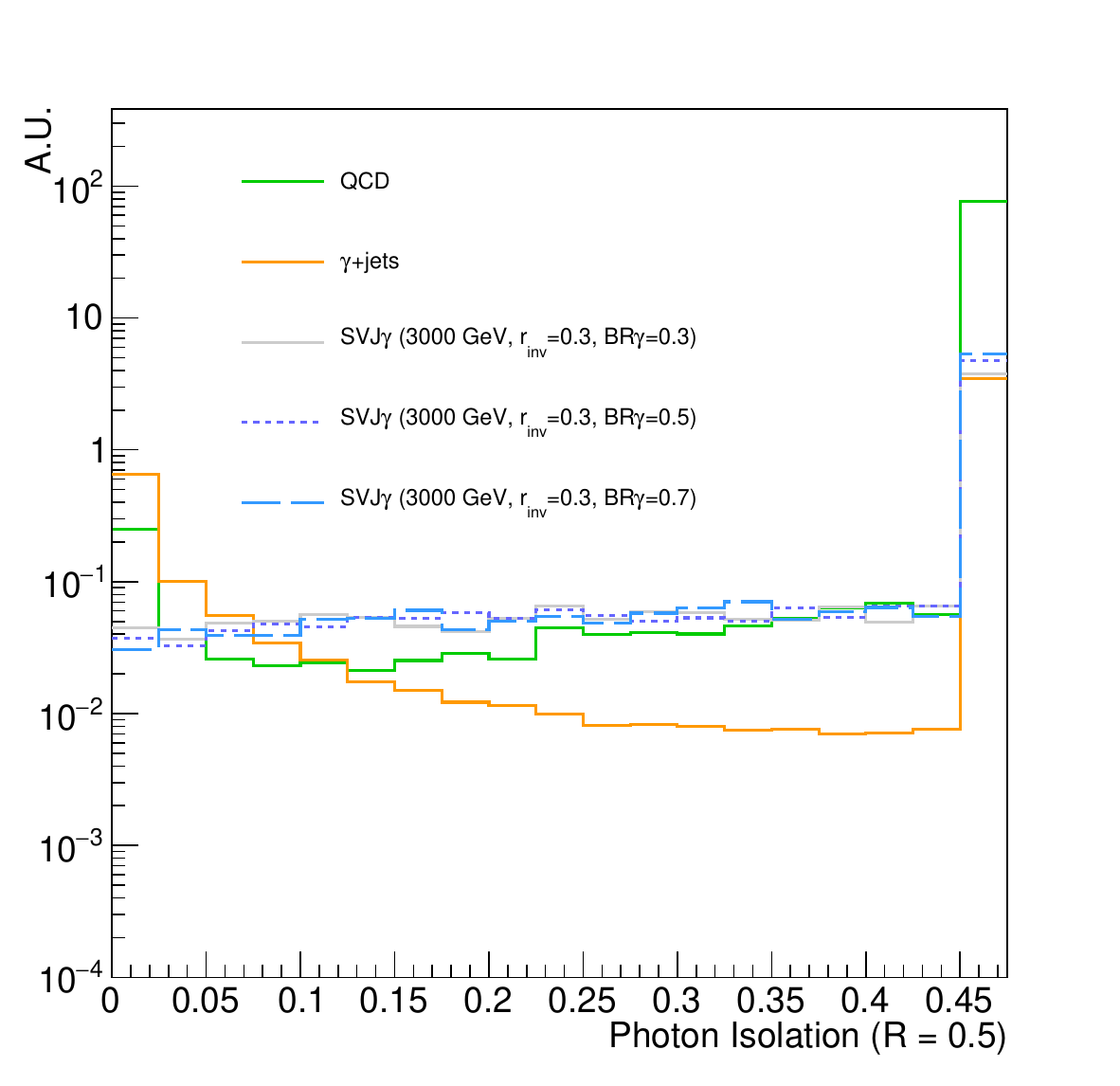}  \label{fig:distr_isolation} 
\end{subfigure} 
\caption{\textbf{(left)} a) Distribution of the neutral electromagnetic fraction (NEMF) for the two highest $p_\mathrm{T}$ AK8 jets for different signal benchmarks and QCD, $\gamma+\mathrm{jets}$ backgrounds. \textbf{(right)} b) Distribution of photons isolation for~$R_{\text{iso}}= 0.5$ for different signal benchmarks and QCD, $\gamma+\mathrm{jets}$ backgrounds. The last bin represents the overflow. The signal distributions are referred to a benchmark mass point~$M_{Z'}= \SI{3}{TeV}$, $r_{\text{inv}} = 0.3$ and $\text{BR}_{\gamma} = 0.3, \ 0.5, \ 0.7$.}
\label{fig:distributions_id} 
\end{figure*}

\section{Search strategies}
\label{sec:3}

In this section, we show how the analysis strategy proposed in published experimental results from resonant hadronic SVJ searches~\cite{CMS:2021dzg} is not sensitive to the signature proposed in this Letter. Indeed, one would expect major constraints on SVJ$\gamma$ signature from such results. In the following, we report the object and event selections of~\cite{CMS:2021dzg}, applied to SVJ$\gamma$ events to verify sensitivity of available searches:
\begin{itemize}
    \item Number of good jets: $N(j^\text{AK8})\geq2$, $p_\text{T}(j_{1,2}^\text{AK8})>\SI{200}{GeV}$, $|\eta(j_{1,2}^\text{AK8})|<2.4$ and passing identification criteria (JetID)
    \item Jet separation: $\Delta\eta(j_1^\text{AK8},j_2^\text{AK8})<1.5$
    \item Transverse mass selection~\footnote{The di-jet transverse mass is computed from the 4-vector of the di-jet system $(E_{\text{T},jj},\vec{p}_{\text{T},jj})$ and $(\cancel{E}_{\text{T}},\vec{\cancel{E}_{\text{T}}})$ \cite{Cohen:2015toa}: $M^2_{\text{T}} = (E_{\text{T},jj}+\cancel{E}_{\text{T}})^2 - (\vec{p}_{\text{T},jj}+\vec{\cancel{E}_{\text{T}}})^2$. The requirement on the transverse mass allows to select events for which the Run2 jet HT-based triggers employed in~\cite{CMS:2021dzg} are fully efficient.}: $M_{\text{T}}(j_1^\text{AK8},j_2^\text{AK8})>\SI{1.5}{TeV}$ 
    \item Jet-$\cancel{E}_{\text{T}}$ separation: $\Delta\phi_{\text{min}}(\cancel{E}_{\text{T}},~j_{1,2}^\text{AK8}) < 0.8$
\end{itemize}
\noindent
The jet identification criteria (JetID) are usually applied to select good jets that can be used for analysis~\cite{CMS-DP-2024-028}. It consists of a set of cuts which
aims to reject fake, badly
reconstructed and noise jets while retaining a large part of the real jets (around \SIrange{98}{99}{\%}). The JetID employed by ATLAS and CMS experiments generally makes use of jet energy fractions and the number of jet constituents to select high-quality jets. Namely, to reject photons resembling jets, typically the fraction of energy in a jet carried by photons (neutral electromagnetic fraction, NEMF) is required to be less than 0.9 (0.99) for $|\eta| < 2.6$ ($2.6<|\eta| < 2.7$)~\cite{CMS-DP-2024-028}. Very high values of NEMF point to energy deposits from photons.  Figure~\ref{fig:distributions_id}\textcolor{blue}{.a} shows the expected NEMF distribution for different signals and backgrounds. A clear limitation of analysis employing jets as final state objects in terms of sensitivity to the SVJ$\gamma$ signature comes from such a requirement on the NEMF, which rejects most of the signal. Considering a reference mass point at $M_{Z'}=\SI{3}{TeV}$, the signal efficiency decreases from around \SIrange{14}{20}{\%} (see Table~\ref{tab:cut_flows_s_with_nemf}, \ref{app:cut_flows}) down to \SI{0.4}{\%} when applying the NEMF selection (see Table~\ref{tab:cut_flows_s_with_nemf}, \ref{app:cut_flows}). Removing the requirement on the NEMF from the JetID implies higher signal efficiencies while keeping the background efficiencies the same (see Table~\ref{tab:cut_flows_b_no_nemf} and~\ref{tab:cut_flows_b_nemf}). Thus a signal significance improvement is expected (look at section~\ref{sec:5}).\\
New physics searches looking for jets final states, like the CMS resonant SVJ search, are not the only ones that might be sensitive to  SVJ$\gamma$. Analyses focusing on photons final states, such as the high-mass di-photon searches~\cite{CMS:2016xbb}, might show some sensitivity. However, the efficiency of SM photon selections to the SVJ$\gamma$ signature is expected to be very limited. Indeed, photons in SVJ$\gamma$ are expected to be non-isolated due to nearby hadrons coming from the unstable dark bound states decays. Figure~\ref{fig:distributions_id}\textcolor{blue}{.b} shows how the majority of the signal photons lie at high values of the energy deposits around the photon (\emph{isolation}), while most of the background from prompt photons remains at very low values. Typically, analysis exploiting photons requires isolation (or equivalently, the ratio between the energy deposited in the hadronic and electromagnetic calorimeters) to be small, thus removing the majority of our signal. 

From our studies is clear how SM jets and photon identification criteria are suboptimal for identifying SVJ$\gamma$, pointing to the need for a dedicated identification strategy. 

\begin{figure} 
\centering 
\includegraphics[width=.99\linewidth]{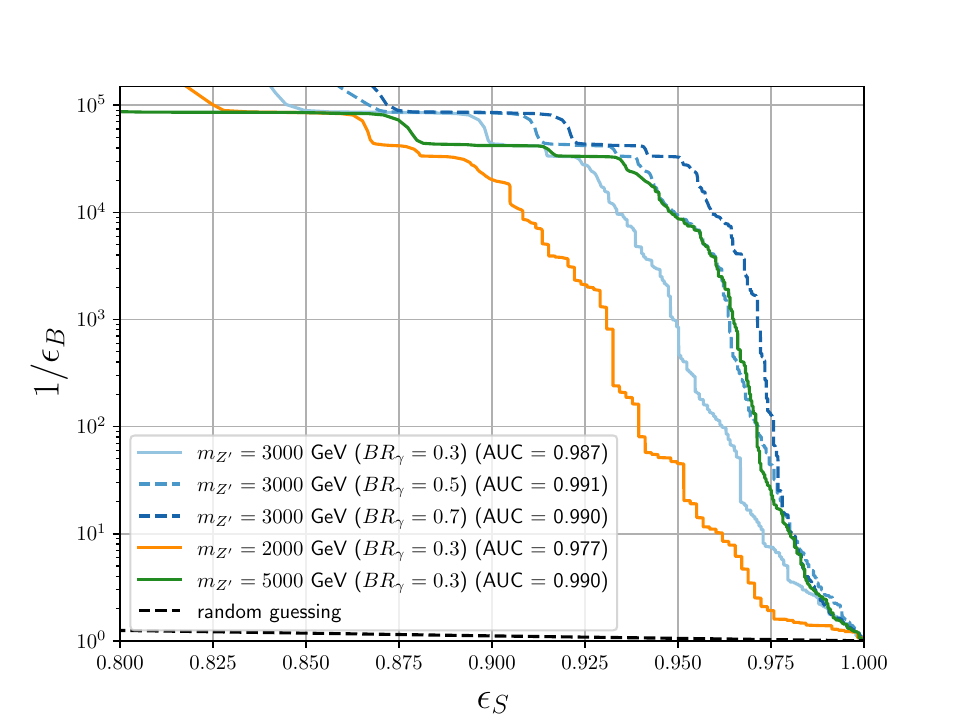} \label{fig:roc_plot} 
\caption{Performance of the DNN-based jet tagger scanning on $M_{Z'} \in [2,3,5]$~TeV with fixed $\text{BR}_{\gamma} = 0.3$ (solid lines), and scanning over $\text{BR}_{\gamma} \in [0.3,0.5,0.7]$ with fixed $M_{Z'} = \SI{3}{TeV}$. The signal efficiency is reported on the $x$ axis, while the reciprocal of the background efficiency is reported on the y axis. The areas under the receiver operating characteristic curves (curves in $\epsilon_B-\epsilon_S$ plane) are reported in the legend. The dashed black line represents random guessing, and all the performance curves are far above this line, thus showing high signal-over-background discrimination.\label{fig:DNN_performance} } 
\end{figure}

\section{SVJ$\gamma$ identification}
\label{sec:4}
The selection requirements employed by the CMS SVJs search \cite{CMS:2021dzg} rely on event-level variables (including missing transverse momentum)  or basic kinematic properties of the reconstructed jets and missing transverse momentum. Even if this selection rejects the vast majority of the SM background events (Table~\ref{tab:cut_flows_b_no_nemf}, \ref{app:cut_flows}), it does not capture the peculiarities of the SVJ$\gamma$ signature since the expected enrichment in photons is not used anywhere to discriminate against the background. Moreover, removing the NEMF criteria from the JetID, even if keeping similar expected background rates, results in retaining more jets which are faked by a photon, especially $\gamma+\text{jets}$ background, as can be seen from the peak near 1 from Figure~\ref{fig:distributions_id}\textcolor{blue}{.a}. In order to reduce the QCD and $\gamma+\text{jets}$ backgrounds further and enhance the sensitivity of the search, we employ the tools developed in the context of jet substructure physics~\cite{Brooijmans:1077731,Butterworth:2007ke,Butterworth:2008iy,Thaler:2011gf,Kaplan:2008ie} to design a novel jet tagging algorithm leveraging the specific features of signal jets.
In our jet tagger, we exploit 3 main categories of jet substructure variables: generalized angularities~\cite{Larkoski:2014pca} and jet shape observables~\cite{CMS:2013kfa}, N-subjettiness~\cite{Thaler:2010tr} and energy fractions. The generalized angularities describe the spatial distribution of the constituents in a jet as well as their momentum flow, and in a similar way, the jet shape observables encode the information about the spatial distribution of the constituents. The N-subjettiness provides a simple way to effectively count the number of subjets inside a given jet. It captures whether the energy flow inside a jet deviates from the one-prong configuration expected to characterize a typical QCD-jet. Finally, the third class of variables includes the energy fractions, which describe the energy carried by specific final states objects (such as electrons, photons, muons and charged hadrons) inside the jet. The list of all 17 input variables used with their definition can be found in~\ref{app:jet_tagger}.  \\ \\
These variables are used as input to a Deep Neural Network (DNN). The DNN is trained using the Pytorch package~\cite{paszke2017automatic}. The inputs are the two highest $p_{\text{T}}$ jets from simulated signal and background samples, with the variables described above computed for each jet. The DNN is trained with each signal sample weighted equally and considering the different backgrounds with the same proportions with respect to one another. More details on the training of the network, the choice of the hyperparameters, and the architecture can be found in~\ref{app:dnn_training}. \\
The DNN exhibits strong rejection of jets from the SM background processes. The performance curves of the tagger as a function of model parameters versus all background jets are shown in Figure~\ref{fig:DNN_performance}. The three most important variables exploited by the DNN to discriminate the signal from the total background, evaluated using SHapley Additive exPlanations (SHAP)~\cite{trumbelj2014ExplainingPM,10.5555/3295222.3295230} with the shap python package~\cite{NIPS2017_7062,lundberg2020local2global,mitchell2022gputreeshapmassivelyparallelexact,lundberg2018explainable}, have been found to be the transverse momentum dispersion ($p_{T}D$), the photon energy fraction ($f_{\gamma}$) and the 4-subjettiness ($\tau_4$). The distributions of some of the most relevant DNN input features are shown in Figure \ref{fig:dnn_most_important_input_vars_plots} (\ref{app:dnn_input_features}). In particular, we observe that the signal usually reaches higher values for $p_{T}D$ compared to QCD, $t \bar{t}+\text{jets}$ and $Z/W+\text{jets}$, meaning that most of the momentum is carried by few of the signal jets constituents. While, for the $\gamma+\text{jets}$ background, since some jets are just made up of one photon, the $p_{T}D$ distribution is expected to peak at 1 and can still be separated from the signal which has a broader spectrum. Regarding the $f_{\gamma}$ distribution, as mentioned in section~\ref{sec:3}, the signal is expected to be distributed at larger values compared to QCD, $t \bar{t}+\text{jets}$ and $Z/W+\text{jets}$ because of the enriched content in photons, while for $\gamma+\text{jets}$ the distribution is expected to peak at 1 for a fraction of the jets, however, the shape of the signal is expected to be different as shown in Figure~\ref{fig:distributions_id}\textcolor{blue}{.a}. Finally, the signal is expected to reach smaller values for $\tau_4$ due to the multi-prong substructure, which characterises it with respect to all SM backgrounds. \\
We choose a working point (WP) corresponding to a threshold of 0.8 on the discriminator output. Jets with a discriminator value higher than 0.8 are labeled as semi-visible. The distributions of the DNN score for signal and background with the chosen selection are reported in Figure \ref{fig:DNN-score} (\ref{app:jet_tagger}). At this WP, the DNN rejects $\sim \SI{99.6}{\%}$ of the simulated total background jets, while correctly classifying $\sim \SI{95.2}{\%}, \SI{96.6}{\%}, \SI{97.3}{\%}$ of jets from the benchmark signal models at $M_{Z'}=\SI{3}{TeV}$ for $\text{BR}_{\gamma}=0.3, 0.5, 0.7$, respectively. In particular, for the chosen WP, the DNN leaves $\SI{0.23}{\%}$ of QCD jets, $\SI{0.5}{\%}$ of $t\bar{t}$ jets, $\SI{0.6}{\%}$ of jets from $\gamma+\mathrm{jets}$ and $\SI{0.9}{\%}$ of jets from electroweak processes. 
\section{Results}
\label{sec:5}
\begin{figure}[!htbp]
  \centering
  \includegraphics[width=0.53\textwidth]{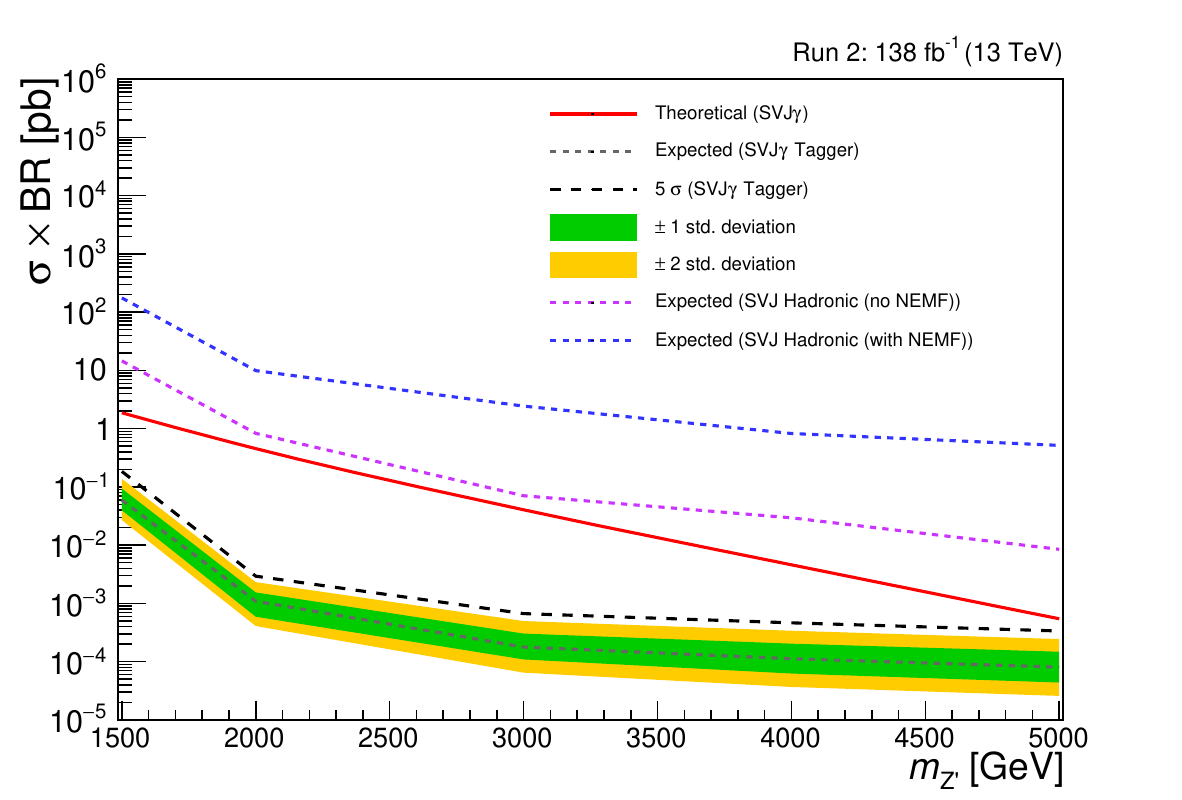}
\caption{Expected limits for Run~2 on $\sigma \times BR(Z' \to q_v \bar{q}_v)$ for the signal benchmark with $r_{\text{inv}} = 0.3$ and $\text{BR}_{\gamma} = 0.3$. The expected exclusion reach for the SVJ$\gamma$ signature from the fully hadronic SVJ inclusive analysis (applying the NEMF requirement in the JetID) (dashed blue line) is compared with both the cut-based strategy removing the NEMF selection (purple dashed line) and the SVJ$\gamma$ tagger-based analysis (grey dashed line), exploiting the jet tagger introduced in section \ref{sec:4}. The black dashed line line corresponds to the expected $5\sigma$ discovery. The red line represents the theoretical prediction computed assuming the coupling of $Z'$ to SM quarks to be 0.25 and the coupling to dark quarks to be 0.4.}
\label{fig:limit_result}
\end{figure}
\noindent
We report in this section the sensitivity and exclusion reach for the SVJ$\gamma$ signal. We have estimated the expected exclusion limit at 95\% confidence level (CL) for $\sigma \times \text{BR}$ for different $M_{Z'}$ and $\text{BR}_{\tau}$ hypotheses performing a binned likelihood template fit of the $M_{\text{T}}$ spectrum and using the modified frequentist approach $\text{CL}_{\text{s}}$ in the asymptotic approximation~\cite{Junk:1999kv,Read:2002hq,Cowan:2010js}. 
The sensitivity reach has been estimated for events selected according to the strategy defined in section~\ref{sec:4}. We have combined three categories of events according to the jet-level decision of the tagger at chosen optimal WP (as introduced in section~\ref{sec:5}): 0-tag category, where no jets are tagged as semi-visible, 1-tag category where one jet is tagged as semi-visible and 2-tag category where one jet is tagged as semi-visible. \\
We report the results for a benchmark point with $r_{\text{inv}}= 0.3$ and $\text{BR}_{\gamma} = 0.3$. The LHC with full Run~2 collision data ($\sqrt{s}=\SI{13}{TeV}$, $\mathcal{L}_{int} = \SI{138}{{fb}^{-1}}$) and following the tagger-based approach proposed here is expected to claim the discovery (exclusion) of a hypothetical $Z'$ boson in the full mass range considered for a benchmark coupling of $Z'$ to SM quarks equal to 0.25, and the coupling to dark quarks equal to 0.4. More details are shown in Figure~\ref{fig:limit_result}. We observe that the sensitivity to the signal slightly increases with higher values of $\text{BR}_{\gamma}$ due to  the higher signal efficiency and tagger discrimination power. Furthermore,  Figure \ref{fig:limit_result}\textcolor{blue}, shows how the inclusive CMS SVJ search~\cite{CMS:2021dzg} has no sensitivity to SVJ$\gamma$ signals, while the jet tagger proposed in this Letter allows probing values of the coupling of the $Z'$ portal to SM quarks even smaller than the benchmark 0.25, for the entire $M_{Z'}$ mass range tested.\\

\section{Conclusions}
\label{sec:6}
This Letter proposes a novel experimental signature and relevant search strategy for the discovery of confining hidden sectors that give rise to SVJs with non-isolated prompt photon pairs. We have introduced a simplified model based on two messenger bosons, namely a $Z'$ and an ALP, allowing the resonant production of SVJs enriched in photons. 
The phase space of the SVJ$\gamma$ signal is expected to be loosely constrained by the existing searches due to the tight NEMF and isolation requirements used in the standard jets and photons identification criteria. We have shown that the fully hadronic SVJ s-channel search has poor sensitivity to the SVJ$\gamma$ signature, and removing (or loosening) the NEMF requirement might not be enough to get the necessary sensitivity to discover the signal. Thus, we developed a DNN-based jet tagging algorithm based on jet substructure features to enhance the signal-to-background separation and boost the sensitivity of the proposed analysis strategy. The proposed search can claim a discovery (exclusion) of the $Z'$ mediator in the full mass range considered with \mbox{$\mathcal{L}_{int} =\SI{138}{{fb}^{-1}}$}, if the jet tagging algorithm proposed here is applied. \\
Finally, the model proposed in this Letter also allows for displaced signatures in cases of light-dark-bound states. This would lead to signatures with additional displaced tracks and vertices inside the signal jets (similar to emerging jets~\cite{Schwaller_2015}) and non-prompt, non-isolated photons. This signature requires a dedicated strategy that is not addressed in this Letter, but might be of interest for future works.

\begin{acknowledgements}
 C. Cazzaniga and A. de Cosa  are supported by the Swiss National Science Fundation (SNFS) under the SNSF Eccellenza program ($\mathrm{PCEFP2}\_\mathrm{186878}$).
\end{acknowledgements}

\appendix

\section{Event selection}
\label{app:cut_flows}
The event selections used in this work are shown in the first columns of Tables \ref{tab:cut_flows_s_no_nemf} and \ref{tab:cut_flows_b_no_nemf}, respectively for a benchmark signal and the backgrounds. This is based  on a previous CMS study~\cite{CMS:2021dzg}, and it is updated removing the requirement on the NEMF in the JetID. The last three columns show the signal efficiencies for different values of BR$_\gamma$ and a benchmark mass point $M_{Z'}=\SI{3}{TeV}$. In Table~\ref{tab:cut_flows_s_with_nemf} for a benchmark signal and Table \ref{tab:cut_flows_b_nemf} for the backgrounds, the event selection including the NEMF requirement as JetID are shown.

\begin{table}[htb]
	\centering
    \renewcommand{\arraystretch}{1}
	\begin{tabular}{c c c c }\hline 
	\multirow{2}{2em}{Selection} &
	    \multicolumn{3}{c}{Signal efficiency ($\%$)} \\
		$\quad \quad \quad \quad \quad \quad \quad \quad \quad \quad \quad \quad\text{BR}_{\gamma}:$& 0.7 & 0.5 & 0.3 \\\hline \hline
		$N(\text{AK8 jets with NEMF}) \geq 2$           & 78.00    & 69.34    & 52.71  \\\hline
		$\Delta\eta (j_1,j_2)  < 1.5$  &  52.70  &  47.74   & 37.59   \\\hline
		$M_{\text{T}} > \SI{1500}{GeV}$                & 42.66   &   35.62   &  23.84   \\\hline
		$R_{\text{T}}>$  0.15             &   23.38   &   22.12    &  16.79   \\\hline
		$\Delta\phi_{\text{min}}(\cancel{E}_{\text{T}},j) <$  0.8  &   21.04  &   19.43    & 13.98   \\\hline
		\end{tabular}
		\caption{Event selections (without the NEMF requirement in the JetID for AK8 jets) applied to SVJ$\gamma$ signal for a benchmark mass point $M_{Z'}=\SI{3}{TeV}$ and varying the branching into photons $\text{BR}_{\gamma}$.}
		\label{tab:cut_flows_s_no_nemf}
\end{table}
\renewcommand{\arraystretch}{0.2}

\begin{table}[htb]
	\centering
    \renewcommand{\arraystretch}{1}
	\begin{tabular}{c c c c }\hline 
	\multirow{2}{2em}{Selection} &
	    \multicolumn{3}{c}{Signal efficiency ($\%$)} \\
		$\quad \quad \quad \quad \quad \quad \quad \quad \quad \quad \quad \quad\text{BR}_{\gamma}:$& 0.7 & 0.5 & 0.3 \\\hline \hline
		$N(\text{AK8 jets without NEMF}) \geq 2$           & 1.60   &  1.51   & 1.54  \\\hline
		$\Delta\eta (j_1,j_2)  < 1.5$  & 1.00   &  1.05   & 1.08  \\\hline
		$M_{\text{T}} > \SI{1500}{GeV}$                &  0.74  &  0.74    & 0.73    \\\hline
		$R_{\text{T}}>$  0.15             &  0.47    &   0.50     &   0.60  \\\hline
		$\Delta\phi_{\text{min}}(\cancel{E}_{\text{T}},j) <$  0.8  &  0.39   &   0.38    & 0.38  \\\hline
		\end{tabular}
		\caption{Event selections (including the NEMF requirement in the JetID for AK8 jets) applied to SVJ$\gamma$ signal for a benchmark mass point $M_{Z'}=\SI{3}{TeV}$ and varying the branching into photons $\text{BR}_{\gamma}$. The JetID applied here consists only of a requirement on the NEMF (NEMF is required to be less then 0.9 (0.99) for $|\eta| < 2.6$ ($2.6<|\eta| < 2.7$)).}
		\label{tab:cut_flows_s_with_nemf}
\end{table}
\renewcommand{\arraystretch}{0.2}

\begin{table*}[htb]
	\centering
    \renewcommand{\arraystretch}{1}
	\begin{tabular}{c c c c c c}\hline 
	\multirow{2}{2em}{Selection} &
	    \multicolumn{5}{c}{Backgorund efficiency ($\%$)} \\
		$\quad \quad \quad \quad \quad \quad \quad \quad \quad \quad \quad \quad$ & QCD & $\text{t} \bar{\text{t}} + \text{jets}$ & $\text{Z}(\nu \bar{\nu}) + \text{jets} $ & $\text{W} (\ell \nu) + \text{jets}$ & $\gamma + \text{jets}$ \\\hline \hline
		$N(\text{AK8 jets with NEMF}) \geq 2$  & 98.17    &  7.19   &  1.04 & 1.59 & 98.55 \\\hline
		$\Delta\eta (j_1,j_2)  < 1.5$  &  66.53  &  5.33   & 0.67  & 1.11 & 68.40   \\\hline
		$M_{\text{T}} > \SI{1500}{GeV}$                & 14.98   &  0.15    &  0.04   & 0.03  & 17.37   \\\hline
		$R_{\text{T}}>$  0.15             &   0.69   &  0.03    &  0.02  & 0.01  &  0.81 \\\hline
		$\Delta\phi_{\text{min}}(\cancel{E}_{\text{T}},j) <$  0.8  & 0.68    &  0.03 & 0.01  & 0.01 &  0.79  \\\hline
		\end{tabular}
		\caption{Event selections  (without the NEMF requirement in the JetID for AK8 jets) applied to the backgrounds.}
		\label{tab:cut_flows_b_no_nemf}
\end{table*}
\renewcommand{\arraystretch}{0.2}

\begin{table*}[htb]
	\centering
    \renewcommand{\arraystretch}{1}
	\begin{tabular}{c c c c c c}\hline 
	\multirow{2}{2em}{Selection} &
	    \multicolumn{5}{c}{Backgorund efficiency ($\%$)} \\
		$\quad \quad \quad \quad \quad \quad \quad \quad \quad \quad \quad \quad$ & QCD & $\text{t} \bar{\text{t}} + \text{jets}$ & $\text{Z}(\nu \bar{\nu}) + \text{jets} $ & $\text{W} (\ell \nu) + \text{jets}$ & $\gamma + \text{jets}$ \\\hline \hline
		$N(\text{AK8 jets with NEMF}) \geq 2$  &   98.03  &  7.19   & 1.04  & 1.57 & 89.16 \\\hline
		$\Delta\eta (j_1,j_2)  < 1.5$  & 66.44   &  5.33   & 0.67  & 1.10 & 60.86  \\\hline
		$M_{\text{T}} > \SI{1500}{GeV}$                &  14.96   &  0.15    &  0.03  & 0.03 & 15.68   \\\hline
		$R_{\text{T}}>$  0.15             &  0.69     &  0.03    &  0.02  & 0.01 & 0.76  \\\hline
		$\Delta\phi_{\text{min}}(\cancel{E}_{\text{T}},j) <$  0.8  &  0.68   & 0.03   & 0.01  & 0.01 & 0.75  \\\hline
		\end{tabular}
		\caption{Event selections (including the NEMF requirement in the JetID for AK8 jets) applied to the backgrounds. The JetID applied here consists only of a requirement on the NEMF (NEMF is required to be less then 0.9 (0.99) for $\eta < 2.6$ ($2.6<\eta < 2.7$)).}
		\label{tab:cut_flows_b_nemf}
\end{table*}
\renewcommand{\arraystretch}{0.2}

\section{DNN jet tagger input features}\label{app:jet_tagger}
Here we report the definition of the input features used for the DNN training. All input features have been computed using AK8 jets and their constituents. The first category of variables we used are the generalized angularities, and they are defined from the constituents $i \in \{1,\cdots ,N\}$ carrying momentum fraction $z_i$ inside a jet of cone size $R$ as:
\begin{equation}\label{eq:generalized_angularities}
\lambda^{\kappa}_{\beta} = \sum_{i} z^{\kappa}_{i} \biggr( \frac{\Delta R_{i,\text{jet}}}{R} \biggr)^{\beta}
\end{equation}
where the sum is over the jet constituents. From Eq.~\ref{eq:generalized_angularities}, the input features we used for the training of the network are the following:
\begin{itemize}
    \item girth ($\lambda^{2}_{0}$) 
    \item LHA ($\lambda^{1}_{\frac{1}{2}}$)
    \item thrust ($\lambda^{1}_{2}$) 
    \item transverse momentum dispersion ($p_{T}D = \lambda^{2}_{0}$) 
    \item multiplicity ($\lambda^{0}_{0}$)
\end{itemize}
The first three features describe the spatial distribution of the constituents in jets, while $p_{T}D$ encodes how the momentum is distributed between the constituents of the jet. \\ \\
We supplement such set of features with further jet shape variables known as axes. The shape of the jet can be approximated by an ellipse in the $\eta-\phi$ plane. The major and minor axes are the two principal components of this ellipse and are defined from the following symmetric matrix $M$:
\begin{equation}\label{eq:jet_shape_matrix}
M = \begin{bmatrix}
\sum_i p^2_{T,i}(\Delta \eta_i)^2 & -\sum_i p^2_{T,i}\Delta \eta_i\Delta \phi_i \\
-\sum_i p^2_{T,i}\Delta \eta_i\Delta \phi_i & \sum_i p^2_{T,i}(\Delta \phi_i)^2 
\end{bmatrix}
\end{equation}
where the sum runs over all constituents of the jet and $\Delta \eta$, $\Delta \phi$ are the differences in $\eta$ and $\phi$ with respect to the jet axis. The major and minor axes are defined from the eigenvalues $\lambda_1$ and $\lambda_2$ of $M$ as:
\begin{equation}\label{eq:axes_major_minor}
\sigma_{\text{minor,major}} = \sqrt{\frac{\lambda_{1,2}}{\sum_i p^2_{T,i}}}
\end{equation}
The average axis can be defined from $\sigma_{\text{minor,major}}$ as: 
\begin{equation}\label{eq:axes_avg}
\sigma_{\text{avg}} = \sqrt{\sigma^2_{\text{minor}} + \sigma^2_{\text{major}} }
\end{equation}
The second set of features we feed into the DNN are the N-subjettiness~\cite{Thaler:2010tr}. The N-subjettiness $\tau^{\beta}_N$ are designed to count the number of subjets inside a jet. In specific, N-subjettiness is defined as:
\begin{equation}\label{eq:n_subjettiness}
\tau^{\beta}_N = \sum_i p_{T,i}\text{min} (R^{\beta}_{1,i},R^{\beta}_{2,i}, \cdots, R^{\beta}_{N,i})
\end{equation}
where the sum is over the jet constituents, and $R^{\beta}_{N,i}$ is the distance between the $N$th subjet
and the $i$th constituent of the jet. $\tau^{\beta}_N$ measures departure from $N$-parton energy flow: if a jet has $N$ subjets, $\tau^{\beta}_N$ should be much larger than $\tau^{\beta}_N$. For our DNN training we use N-subjettiness with $N \in \{1,\cdots,5\}$ and $\beta = 1$. \\ \\
Finally, the third set of variables we used are energy fractions. These quantities can be defined for different jet constituents, such as electrons ($e$), muons ($\mu$), photons ($\gamma$) and charged hadrons ($h_{ch}$). Fixed a specific final state object $\tilde{c}$ among those listed above, the energy fraction relative to that object is defined as:
\begin{equation}\label{eq:energy_fractions}
f_{\tilde{c}} = \sum_i \frac{E_{\tilde{c},i}}{E_i} \quad \quad \tilde{c} \in \{e, \mu, \gamma, h_{\text{ch}} \}
\end{equation}
For the training of the DNN tagger, we use as input features the following energy fractions: $f_{e}, \ f_{\mu}, \ f_{\gamma}, \ f_{h_{\text{ch}}}$.  

\section{DNN architecture and training}\label{app:dnn_training}
The Deep Neural Network has been implemented using the Pytorch package~\cite{paszke2017automatic}. The DNN takes 17 input features, and consists of 2 hidden {\tt linear layers} with 34 nodes each. Each linear layer is followed by a {\tt batch normalization} layer and an activation function. The {\tt relu} activation function has been chosen for the two hidden layers, and the {\tt sigmoid} for the ouput layer of the network. For the output layer, between the batch normalization layer and the sigmoid activation function, a {\tt droput} layer has been added in order to prevent overfitting. The value of the dropout has been chosen to be 0.2. The DNN has been trained to reach converge for 300 epochs with a batch size of 256 and a learning rate of 0.001. The {\tt Adam} optimizer has been employed for the gradient descent.

\section{DNN performance}\label{app:jet_tagger}
\begin{figure}[htbp!]
 \centering
 \includegraphics[width=.4\textwidth]{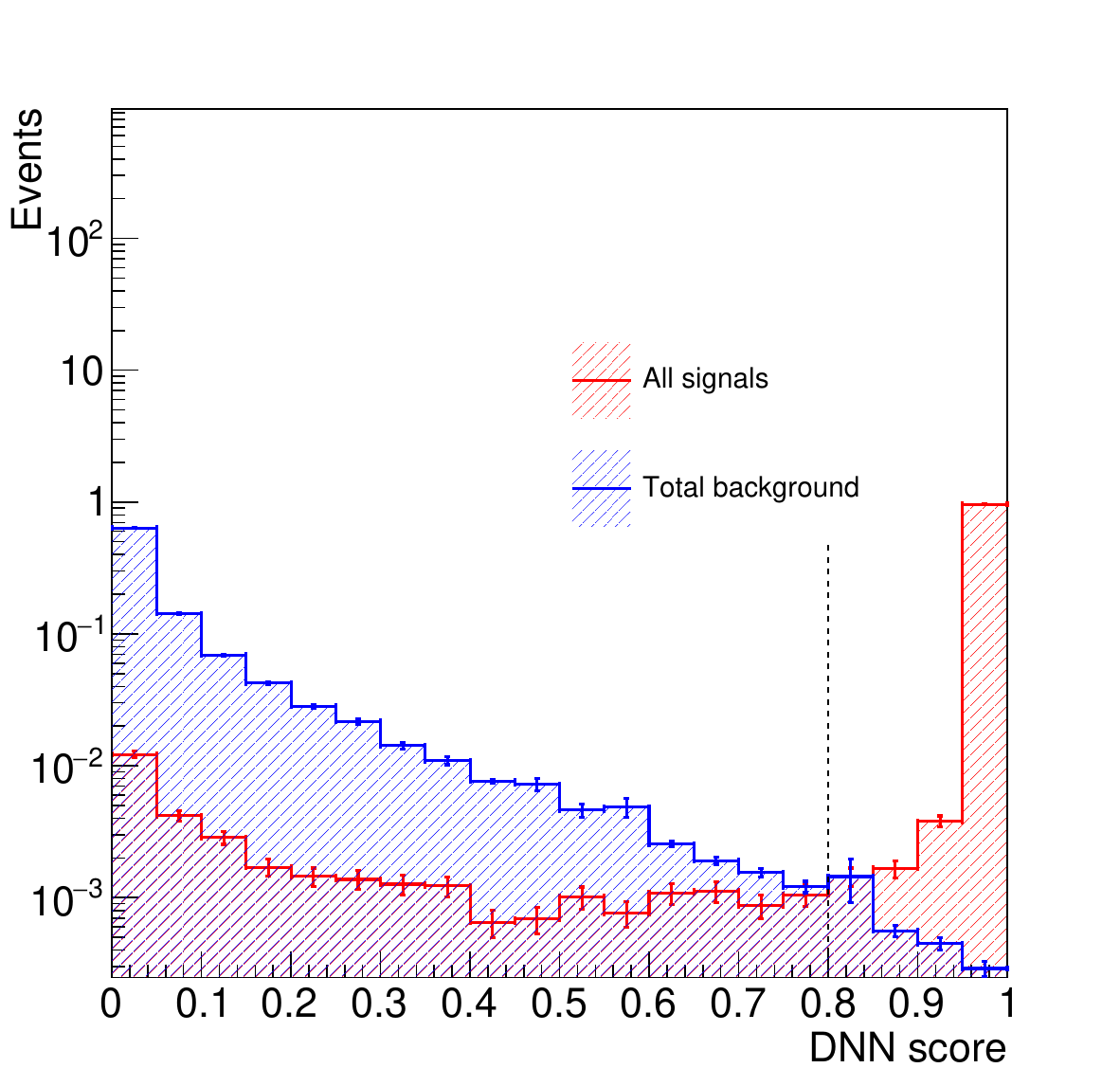}
 \caption{
  \label{fig:DNN-score}
 DNN score distribution for signal (red) and background (blue) test datasets. The vertical black dashed line represents the cut value chosen to define a background jet ($\text{DNN score} < 0.8$) against a signal jet ($\text{DNN score} > 0.8$).    
 }
\end{figure}

The distributions of the DNN scores for signal and background on the test dataset together with the working point chosen is shown in Figure~\ref{fig:DNN-score} .

\section{DNN input features}\label{app:dnn_input_features}
\begin{figure*}[htbp!]
\centering
\includegraphics[scale=.4]{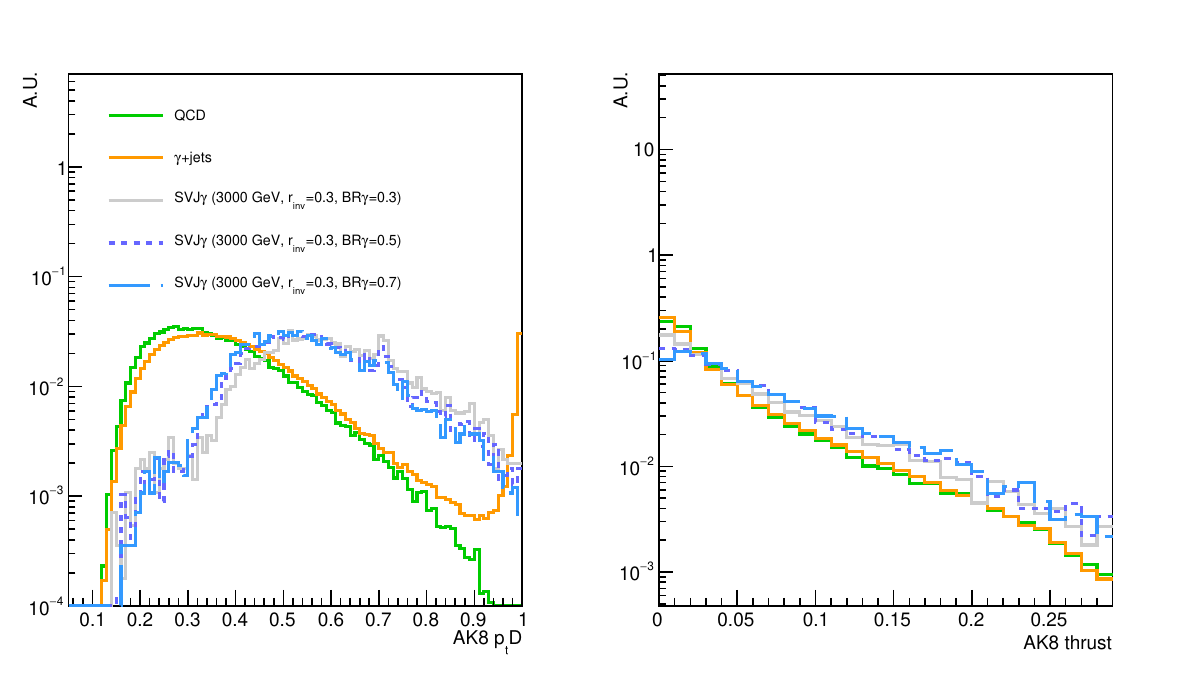}
\includegraphics[scale=.4]{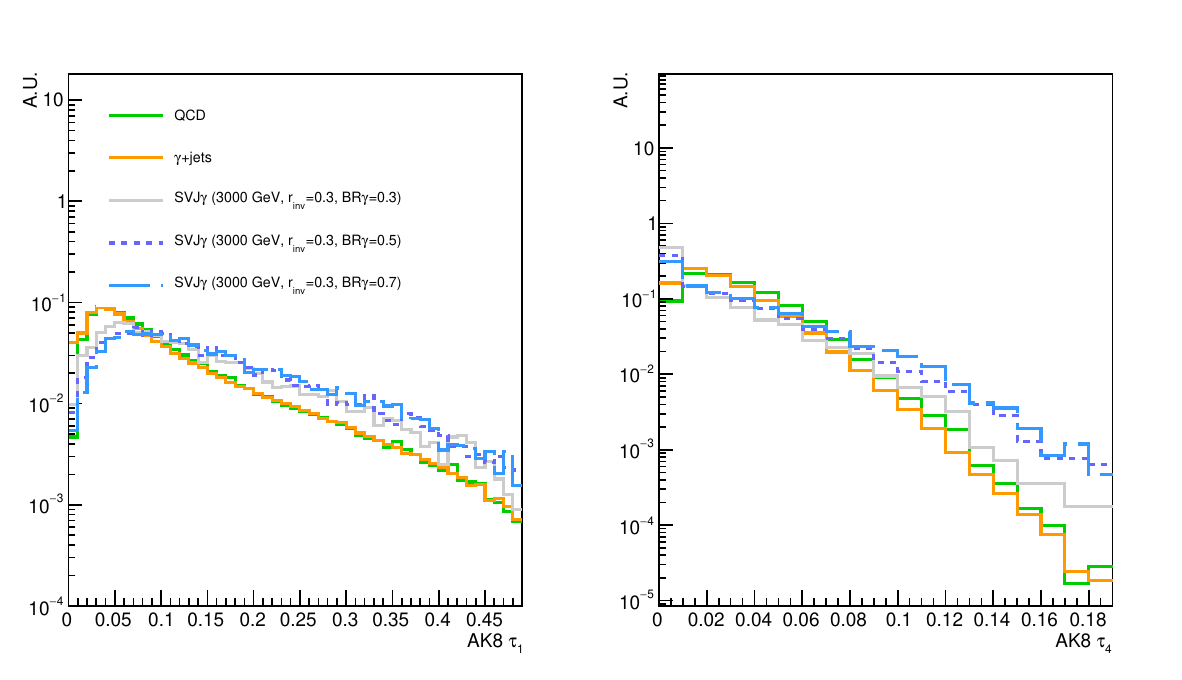}
\includegraphics[scale=.4]{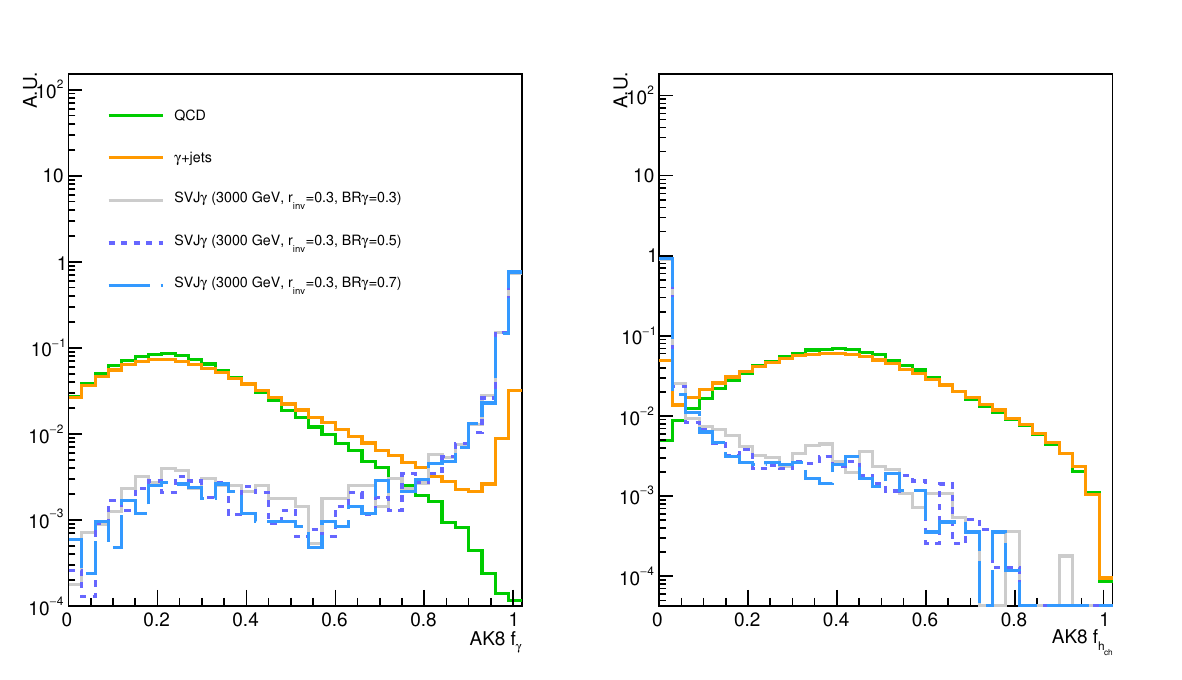}
{\caption{\textbf{(top left)} Distribution of the two most discriminating generalized angularities: $p_{\mathrm{T}}D$ and thrust. \textbf{(top right)} Distribution of the two most discriminating N-subjettiness $\tau_1$ and $\tau_4$. \textbf{(bottom)} Distribution of the two most discriminating energy fractions: $f_{\gamma}$ and $f_{h_{ch}}$. The signal distributions are referred to a benchmark mass point~$M_{Z'}= \SI{3}{TeV}$, $r_{\text{inv}} = 0.3$ and $\text{BR}_{\gamma} = 0.3, \ 0.5, \ 0.7$. The backgrounds plotted are $\gamma+\text{jets}$ (orange line) and QCD (green line), while all the signals are merged. All distributions shown are normalised to one. \label{fig:dnn_most_important_input_vars_plots}}}
\end{figure*}

The distributions of the two most discriminating input features based on SHAP used to train the DNN for each of the three classes are shown in Figure~\ref{fig:dnn_most_important_input_vars_plots} .

\bibliographystyle{spphys}       
\bibliography{bibl}   

\end{document}